\documentclass[prd,a4paper,twocolumn]{revtex4}
\usepackage{graphicx}
\usepackage{subfigure,amssymb,latexsym}
\usepackage{psfrag}

\newcommand{\nc}{\newcommand}

\nc{\be}[1]{\begin{equation}\mbox{$\label{#1}$}}
\nc{\bea}[1]{\begin{eqnarray} \mbox{$\label{#1}$}}
\nc{\Section}[2]{\section{#2}\label{#1}}
\nc{\Bibitem}[1]{\bibitem{#1}}
\nc{\Label}[1]{\label{#1}}

\nc{\eea}{\end{eqnarray}}
\nc{\ee}{\end{equation}}

\nc{\bdm}{\begin{displaymath}}
\nc{\edm}{\end{displaymath}}
\nc{\dpsty}{\displaystyle}
\nc{\bc}{\begin{center}}
\nc{\ec}{\end{center}}
\nc{\ba}{\begin{array}}
\nc{\ea}{\end{array}}
\nc{\bab}{\begin{abstract}}
\nc{\eab}{\end{abstract}}
\nc{\btab}{\begin{tabular}}
\nc{\etab}{\end{tabular}}
\nc{\bit}{\begin{itemize}}
\nc{\eit}{\end{itemize}}
\nc{\ben}{\begin{enumerate}}
\nc{\een}{\end{enumerate}}
\nc{\bfig}{\begin{figure}}
\nc{\efig}{\end{figure}}

\nc{\arreq}{&\!=\!&}
\nc{\arrmi}{&\!-\!&}
\nc{\arrpl}{&\!+\!&}
\nc{\arrap}{&\!\!\!\approx\!\!\!&}
\nc{\non}{\nonumber}
\nc{\align}{\!\!\!\!\!\!\!\!&&}

\def\lsim{\; \raise0.3ex\hbox{$<$\kern-0.75em
      \raise-1.1ex\hbox{$\sim$}}\; }
\def\gsim{\; \raise0.3ex\hbox{$>$\kern-0.75em
      \raise-1.1ex\hbox{$\sim$}}\; }

\nc{\DOT}{\hspace{-0.08in}{\bf .}\hspace{0.1in}}
\nc{\Laada}{\hbox {$\sqcap$ \kern -1em $\sqcup$}}
\nc\loota{{\scriptstyle\sqcap\kern-0.55em\hbox{$\scriptstyle\sqcup$}}}
\nc\Loota{{\sqcap\kern-0.65em\hbox{$\sqcup$}}}
\nc\laada{\Loota}
\nc{\qed}{\hskip 3em \hbox{\BOX} \vskip 2ex}

\nc{\real}{{\rm I \! R}}
\nc{\Z}{{\sf Z \!\!\! Z}}
\nc{\complex}{{\rm C\!\!\! {\sf I}\,\,}}
\def\bigid{\leavevmode\hbox{\small1\kern-3.8pt\normalsize1}}
\def\id{\leavevmode\hbox{\small1\kern-3.3pt\normalsize1}}
\nc{\slask}{\!\!\!/}
\nc{\bis}{{\prime\prime}}
\nc{\pa}{\partial}
\nc{\na}{\nabla}
\nc{\ra}{\rangle}
\nc{\la}{\langle}
\nc{\goto}{\to}
\nc{\swap}{\leftrightarrow}

\nc{\EE}[1]{ \mbox{$\cdot10^{#1}$} }
\nc{\abs}[1]{\left|#1\right|}
\nc{\at}[2]{\left.#1\right|_{#2}}
\nc{\norm}[1]{\|#1\|}
\nc{\abscut}[2]{\Abs{#1}_{\scriptscriptstyle#2}}
\nc{\vek}[1]{{\rm\bf #1}}
\nc{\integral}[2]{\int\limits_{#1}^{#2}}
\nc{\inv}[1]{\frac{1}{#1}}
\nc{\dd}[2]{{{\partial #1}\over{\partial #2}}}
\nc{\ddd}[2]{{{{\partial}^2 #1}\over{\partial {#2}^2}}}
\nc{\dddd}[3]{{{{\partial}^2 #1}\over
    {\partial #2 \partial #3}}}
\nc{\dder}[2]{{{d #1}\over{d #2}}}
\nc{\ddder}[2]{{{d^2 #1}\over{d {#2}^2}}}
\nc{\dddder}[3]{{d^2 #1}\over
    {d #2 d #3}}
\nc{\dx}[1]{d\,^{#1}x}
\nc{\dy}[1]{d\,^{#1}y}
\nc{\dz}[1]{d\,^{#1}z}
\nc{\dl}[1]{\frac{d\,^{#1}l}{(2\pi)^{#1}}}
\nc{\dk}[1]{\frac{d\,^{#1}k}{(2\pi)^{#1}}}
\nc{\dq}[1]{\frac{d\,^{#1}q}{(2\pi)^{#1}}}

\nc{\bfT}{{\bf T }}

\nc{\cA}{{\cal A}}
\nc{\cB}{{\cal B}}
\nc{\cD}{{\cal D}}
\nc{\cE}{{\cal E}}
\nc{\cG}{{\cal G}}
\nc{\cH}{{\cal H}}
\nc{\cL}{{\cal L}}
\nc{\cO}{{\cal O}}
\nc{\cT}{{\cal T}}
\nc{\cN}{{\cal N}}
\nc{\cR}{{\cal R}}
\nc{\rvac}[1]{|{\cal O}#1\rangle}
\nc{\lvac}[1]{\langle{\cal O}#1|}
\nc{\rvacb}[1]{|{\cal O}_\beta #1\rangle}
\nc{\lvacb}[1]{\langle{\cal O}_\beta #1 |}
\nc{\bb}{\bar{\beta}}
\nc{\bt}{\tilde{\beta}}
\nc{\ctH}{\tilde{\cal H}}
\nc{\chH}{\hat{\cal H}}
\nc{\1}{\aa}
\nc{\2}{\"{a}}
\nc{\3}{\"{o}}
\nc{\4}{\AA}
\nc{\5}{\"{A}}
\nc{\6}{\"{O}}
\nc{\al}{\alpha}
\nc{\g}{\gamma}
\nc{\Del}{\Delta}
\nc{\e}{\textrm{e}}
\nc{\eps}{\epsilon}
\nc{\lam}{\lambda}
\nc{\Om}{\Omega}
\nc{\ve}{\varepsilon}
\nc{\mn}{{\mu\nu}}
\nc{\vp}{\varphi}

\nc{\rf}[1]{(\ref{#1})}
\nc{\nn}{\nonumber \\*}
\nc{\bfB}{\bf{B}}
\nc{\bfv}{\bf{v}}
\nc{\bfx}{\bf{x}}
\nc{\bfy}{\bf{y}}
\nc{\vx}{\vec{x}}
\nc{\vy}{\vec{y}}
\nc{\oB}{\overline{B}}
\nc{\oI}{\overline{I}}
\nc{\oR}{\overline{R}}
\nc{\rar}{\to}
\nc{\ti}{\times}
\nc{\slsh}{\hskip-5pt/}
\nc{\sm}{Standard~Model~}
\nc{\MP}{M_{\rm Pl}}
\nc{\mpl}{M_{\rm Pl}}
\nc{\tp}{t_{\rm Pl}}

\nc{\pmin}{p_{\rm min}}
\nc{\pmax}{p_{\rm max}}
\nc{\fo}{f_0}
\nc{\foi}{f_{0,i}\,}
\nc{\fop}{f_0^P}
\nc{\fou}{f_0^U}

\nc{\eff}{{\rm eff}}
\nc{\MT}{M_{\rm T}}
\nc{\ML}{M_{\rm L}}
\nc{\kk}{\vek{k}}
\nc{\pp}{{\rm p}}
\nc{\pt}{\partial_t}
\nc{\half}{{1\over 2}}
\nc{\w}{\omega}
\nc{\uhat}{\hat{U}_\w}

\nc{\etal}{\mbox{\it et al.}}
\nc{\ie}{{\it i.e. }}
\nc{\eg}{{\it e.g. }}
\nc{\trh}{T_{\rm RH}}
\nc{\ad}{{a'\over a}}
\nc{\bd}{{b'\over b}}
\nc{\Rd}{{R'\over R}}
\nc{\diag}{{\textrm{diag}}}
\nc{\mato}[1]{\tilde{#1}}
\nc{\sech}{\textrm{sech}}
\nc{\I}{\textrm{I}}
\nc{\II}{\textrm{II}}
\nc{\III}{\textrm{III}}
\nc{\vev}[1]{\langle #1 \rangle}
\nc{\hyp}{\,\; F_{1{\hskip -16pt}2}{\hskip 11pt}}
\nc{\brhom}{\overline{\rho}_M}
\nc{\brho}{\overline{\rho}}
\nc{\rhob}{\overline{\rho}}
\nc{\Pb}{\overline{P}}
\nc{\bH}{\overline{H}}
\nc{\ep}{{1+4\eps}}

\nc{\lcdm}{$\Lambda$CDM}

\def\smiley{\hbox{\large$\bigcirc$\hspace{-.80em}%
\raise.2ex\hbox{$\cdot\cdot$}\kern-.61em    
\lower.2ex\hbox{\scriptsize$\smile$}}\ }

\def\frowney{\hbox{\large$\bigcirc$\hspace{-.80em}%
\raise.2ex\hbox{$\cdot\cdot$}\kern-.635em
\lower.2ex\hbox{\scriptsize$\frown$}}\ }

\begin{document}

\title{Stellar configurations in $f(R)$ theories of gravity}

\author{K. Henttunen}
\email{kajohe@utu.fi}
\author{T. Multam\"aki}
\email{tuomul@utu.fi}
\author{I. Vilja}
\email{vilja@utu.fi}
\affiliation{Department of Physics, University of Turku, FIN-20014 Turku, FINLAND}

\date{}

\begin{abstract}
We study stellar configurations and the space-time around them in metric 
$f(R)$ theories of gravity. In particular, we focus on the polytropic model 
of the Sun in the $f(R)=R-\mu^4/R$ model. We show how
the stellar configuration in the $f(R)$ theory can, by appropriate initial 
conditions, be selected to be equal to that described by the Lane-Emden -equation 
and how a simple scaling relation exists between the solutions. We also derive the 
correct solution analytically near the center of the star in $f(R)$ theory.
Previous analytical and numerical results are confirmed, indicating that the 
space-time around the Sun is incompatible with Solar System constraints on the 
properties of gravity. Numerical work shows that stellar configurations, with a 
regular metric at the center, lead to $\gamma_{PPN}\simeq1/2$ outside the star \ie 
the Schwarzschild-de Sitter -space-time is not the correct vacuum solution for 
such configurations. Conversely, by selecting the Schwarzschild-de Sitter -metric
as the outside solution, we find that the stellar configuration is unchanged but 
the metric is irregular at the center. The possibility of constructing a 
$f(R)$ theory compatible with the Solar System experiments and possible new 
constraints arising from the radius-mass -relation of stellar objects is discussed.
\end{abstract}

\maketitle

\section{Introduction}

Current cosmological observations provide strong evidence against a critical 
density matter dominated universe. Observations on supernovae type Ia \cite{snia}, 
cosmic microwave background \cite{cmb} and large scale structure \cite{lss}
all indicate that the expansion of the universe is not proceeding as predicted by 
general relativity (GR), if the universe is homogeneous, spatially flat and 
filled with non-relativistic matter. 
The underlying assumptions have hence been questioned, resulting in the emergence 
of a cosmological concordance model, the \lcdm\ -model, describing 
a flat and homogeneous universe dominated by cold dark matter and dark 
energy in the form of a cosmological constant. Alternatives to the simple 
cosmological constant are, however, numerous (for a review see {\it e.g.} 
\cite{peebles}) and other assumptions, such as \eg homogeneity \cite{Alnes} have 
been questioned.

An alternative route to solving the dark energy problem is to consider modifying 
the underlying theory of gravity \ie relaxing the assumption that general 
relativity is the correct theory of gravity on cosmological scales. A popular 
choice is the class of $f(R)$ gravity models that has received much attention in 
the recent literature (see 
\eg \cite{turner,turner2,allemandi,meng,nojiri3,nojiri2,cappo1,woodard,odintsov} 
and references therein).

In a $f(R)$ gravity model, deviations from general relativity arise by 
generalizing the Einstein-Hilbert -action with an arbitrary function of the 
curvature scalar, $f(R)$. Such modification has to face many challenges that 
general relativity passes, including instabilities \cite{dolgov,soussa,faraoni}, 
solar system constraints(see {\it e.g.}\ \cite{chiba,confprobs,Clifton} and 
references therein) and evolution of large scale perturbations 
\cite{Bean:2006up,Song:2006ej}. 

In particular, the solar system observations offer a good testing ground for any 
modification of general relativity by comparing the Parameterized Post-Newtonian 
(PPN) parameters \cite{damour,magnano, olmo, ppnok} with observations. This 
question has recently been extensively reviewed and discussed by a number of 
authors \cite{Erickcek:2006vf,Chiba2,Jin:2006if, Faulkner:2006ub} in order to 
determine the relevance of the Schwarzschild-de Sitter (SdS)-solution in the solar 
system (recently a new class of models that can evade the Solar System 
constraints has been introduced \cite{Hu}).
The SdS-metric is an exact vacuum solution in a large class of 
$f(R)$-theories of gravity that is in agreement with all solar system observations 
with an appropriate cosmological limit. However, the higher order nature of 
$f(R)$ theories makes the SdS-solution (see {\it eg.} \cite{cognola,Multamaki2}) 
non-unique. This property of $f(R)$ theories also demonstrates itself on a 
cosmological scale, making reconstruction of the form of $f(R)$ from the expansion 
history of the universe non-unique \cite{Multamaki}. 

As a result of the recent discussion, it has become clear that the SdS metric is 
unlikely to be the correct vacuum solution in the Solar System for the $1/R$ 
model. Instead, the PPN Solar System constraints are valid in a limit that 
corresponds to the limit of light effective scalar in the equivalent scalar-tensor 
theory.  This is equivalent to requiring that one can approximate the trace of the 
field equations by Laplace's equation \cite{Chiba2} in the corresponding $f(R)$ 
theory. This result has now also been considered by numerical calculations 
\cite{Kainulainen2007}, where the field equations are integrated numerically from 
the center of a star for a fixed matter distribution.

Relatedly, in a recent work \cite{Multamaki3} we considered perfect static fluid 
sphere solutions in $f(R)$ theories of gravity. Again, the higher order nature of 
the $f(R)$ gravity theories demonstrates itself in that the mass distribution 
alone does not uniquely determine the gravitational theory, unless the boundary 
conditions are fixed. If one imposes the SdS-metric as a boundary condition, one 
finds that the solutions are constrained.

Here we consider these questions by numerical and analytical means. We solve the 
set of field equations both inwards and outwards, \ie by starting from the center 
and the boundary of a star. In contrast to \cite{Kainulainen2007}, we do not fix 
the mass distribution beforehand and for completeness also study configurations 
with non-negligible pressure. Furthermore, we consider mass distributions with the 
SdS-metric as a boundary condition
and discuss appropriate analytical limiting solutions near the origin.


\section{$f(R)$ gravity formalism}

The action for $f(R)$ gravity is  
\be{action}
S = \int{d^4x\,\sqrt{-g}\Big(\frac{1}{16\pi G}f(R)+{\cal{L}}_{m}\Big)}
\ee
and the corresponding field equations derived by variating wrt the metric 
$g_{\mu\nu}$ are
\be{eequs}
F(R) R_{\mu\nu}-\frac 12 f(R) g_{\mu\nu}-(\nabla_\mu\nabla_\nu -g_{\mu\nu}\Box )
F(R)=8 \pi G T_{\mu\nu}.
\ee
Here $T_{\mu\nu}$ is the standard minimally coupled stress-energy tensor and 
$F(R)\equiv df/dR$. Contracting the field equations we get another useful form:
\be{econtra}
F(R)R - 2 f(R) + 3\Box F(R) = 8 \pi G (\rho - 3 p).
\ee
We consider spherically symmetric, static configurations ($p=p(r),\ \rho=\rho(r)$) 
and adopt a metric:
\be{metric}
ds^2=B(r)dt^2-A(r)dr^2-r^2(d\theta^2+\sin^2\theta\, d\phi^2).
\ee
In this metric, the non-trivial component of the continuity equation 
$D_\mu T^{\mu\nu}=0$ reads:
\be{cont}
p'=-\frac{B'}{2 B}(\rho+p),
\ee
where $'\equiv d/dr$. Note that, like in GR, the equation of continuity is 
automatically satisfied \cite{koivisto} and hence no additional information is 
gained on top of the field equations. On the other hand, one can choose the 
continuity equation as one of the equations to be solved instead of using the full 
set of field equations.

\subsection{Equations}

For a given action function $f(R)$, one can in principle take any suitable set of 
modified field equations, Eq. (\ref{eequs}) along with the equation of state, 
$p=p(\rho)$, and solve for $\rho, A,\ B$.  However, this can in practice prove to 
be problematic, since the modified Einstein's  equations are highly non-linear and 
high order differential equations, up to fourth order in $B$ and third 
order in $A$.

To simplify the problem, it is useful to consider $f(R)$ and $F(R)$ as independent
functions of $r$. In order to correctly account for their dependence, one then 
needs 
to supplement the set of field equations with an appropriate additional constraint,
$f=f(F)$, determined by the details of the particular $f(R)$ theory in question. 
In this description
we are able to view $F(r),\ A(r),\ B(r)$ and $\rho$ as the fundamental set 
of unknown functions to be solved.  Note that unlike before, now the equation of 
continuity is not automatically satisfied but is an additional, independent 
differential equation. This is due to the fact that $F(r)$ (and $f(r)$) 
is not given in terms of the scalar curvature but is viewed just as a independent 
function of the radial coordinate $r$.

Thus the modified Einstein equation along with the equation of continuity forms 
the set of independent equations to be solved. These nonlinear equations are, 
however, only second order in $F$ and $B$, and first order in $A$ and $\rho$ 
requiring in total six initial conditions for completely determining the solution. 
This is apparently one less that is needed if one proceeds by solving the field 
equations directly in terms of $A$ and $B$, demonstrating that the higher 
derivatives of the modified field equations only appear in the combinations of the 
derivatives of the scalar curvature $R$.

In this paper, we consider the CDTT-model introduced in \cite{turner} with
\be{cdtt}
f(R)=R-\frac{\mu^4}{R},
\ee
but generalization to more complicated models is straightforward. In this 
particular case, it is easy to see that the equation relating $f$ and $F$ is
\be{struct}
f=\mu^2\frac{2-F}{\sqrt{F-1}}.
\ee
In a more general case, like $f(R)=R-\mu^4/R+ R^2/\beta^2$, the functional relation 
is more involved and in general one may need to resort to numerical means.

As is well known, the CDTT-model has a homogeneous solution of de Sitter -metric 
with constant scalar curvature  $R=R_0 \equiv -\sqrt{3}\,\mu^2$. In order to have 
the desired physically plausible late time behaviour, we therefore set 
$\sqrt{3}\,\mu^2\sim H_0^2$.

Using Eq.\ (\ref{struct}) and defining $F(r)\equiv 4/3+v(r), 
\ n(r)\equiv B'(r)/B(r)$, straightforward algebraic manipulations
result in the following set of equations:
\begin{widetext}
\bea{numequs}
\frac{4n}{3r} + \frac{{n}^2}{3} - 8G\pi A\rho + 
  \frac{nv}{r} + \frac{{n}^2v}{4} + 
  \frac{{\mu}^2A}{3 {\sqrt{\frac 13 + v}}} - 
  \frac{{\mu}^2Av}{2{\sqrt{\frac 13 + v}}} - 
  \frac{nA'}{3A} - \frac{nvA'}{4A} + \frac{2n'}{3} + 
  \frac{vn'}{2} - \frac{2v'}{r} + 
  \frac{A'v'}{2A} - v'' & = & 0\nonumber\\
  \frac{{n}^2}{3} + 8G\pi Ap + \frac{{n}^2v}{4} + 
  \frac{{\mu}^2A}{{3}{\sqrt{\frac 13 + v}}} - 
  \frac{{\mu}^2Av}{2{\sqrt{\frac 13 + v}}} - 
  \frac{4A'}{3rA} - \frac{nA'}{3A} - \frac{vA'}{rA} - 
  \frac{nvA'}{4A} + \frac{2n'}{3} + 
  \frac{vn'}{2} - \frac{2v'}{r}-\frac{nv'}{2} & = & 0\nonumber\\
   \frac{4}{3r^2} - \frac{4A}{3r^2} + \frac{2n}{3r} + 8G\pi Ap + 
  \frac{v}{r^2} - \frac{Av}{r^2} + 
  \frac{nv}{2r} - 
  \frac{{\mu}^2Av}{2{\sqrt{\frac{1}{3} + v}}} + 
  \frac{{\mu}^2A}{3{\sqrt{\frac 13 + v}}} - \frac{2A'}{3rA} - 
  \frac{vA'}{2rA} - \frac{v'}{r} - 
  \frac{nv'}{2} + \frac{A'v'}{2A}-v'' & = & 0\nonumber\\
    p'+\frac{n}{2}(p+\rho) & = & 0\nonumber.
\eea
\end{widetext}
This set of equations is suitable for numerical integration made in this article. 

Note, that the dependence on $B(r)$ dissapears completely from the equations 
(\ref{numequs}), \ie the equations are only first order in 
$n(r)$ reflecting the leftover free time scaling of the metric component $B(r)$.

\subsection{Solution near the origin}

In order to solve the field equations numerically from the center, one also has to 
consider the question of boundary conditions. Clearly, one cannot start the 
numerical integration from the origin $r=0$ due to singularities, but a 
small distance from it. Therefore one must select the initial values such that 
they correspond to a desired and a possible solution: otherwise, one might start 
the calculation from a point in parameter space that is unreachable by any 
solution that starts from the origin. 

First we determine the asymptotically correct starting point by considering 
solutions corresponding to regular metrics at the origin. Thus we require 
$A(0),\ n(0),\ \rho(0)$ and $v(0)$ be finite and $p'(0)=0$. Moreover, the radial 
coordinate can be scaled so that $A(0)=1$ as usual. Expanding around the origin 
and solving the field equations gives, up to leading order:
\bea{origsol}
A(r) & = & 1+\frac{8G\pi \left( 3p_0 + 2\rho_0 \right) + 
         \mu^2/\sqrt{\frac{1}{3} + v_0}}{12 + 9v_0}r^2\nonumber\\
n(r) & = & {\frac{16G\pi \left( 3p_0 + 2\rho_0 \right) 
          {\sqrt{\frac 13 + v_0}} - {\mu}^2\left( 2 + 3v_0 \right)}{3
         {\sqrt{\frac 13 + v_0}}\left( 4 + 3v_0 \right) }}r\nonumber\\
v(r) & = & v_0+\left (  {\frac{4G\pi \left( 3p_0 - \rho_0 \right) }{9} - 
       \frac{{\mu}^2v_0}{6{\sqrt{\frac 13 + v_0}}}}\right ) r^2,
\eea
where the subscript $0$ refers to the values at the origin. Pressure and density 
are constant at the level of the approximation. Regularity of the solution at 
origin leaves three free parameters $v_0$, $p_0$ and $\rho_0$, which completely 
determine the solution and hence the stellar structure.  For a given equation of 
state $p=p(\rho)$ the number of parameters further reduces to two. Using these 
equations allows one to start the numerical integration a small 
distance from the origin while preserving the correct asymptotic behaviour.

When one relaxes the requirement that the metric and $v(r)$ are to be finite at 
the origin, it is possible to find several mathematically plausible solutions at 
the vicinity of the origin with finite density $\rho(0)$ but having singular 
behaviour of metric. As we will see, such solutions arise easily whenever one 
tries to solve the field equations inwards, from the stellar boundary to the 
center.

\subsection{Parametrisation of  density and pressure}

In order to consider more realistic matter distribution than predetermined toy 
models, we consider here polytropic stars \ie stars with an equation of state 
$p=\kappa \rho^{\gamma}$. Here $\kappa$ and $\gamma=1+1/n$ are constants and $n$ is 
often referred to as the polytropic index. Such equations of state are useful in 
studying white dwarfs, neutron stars and can also be used as a simple model of 
main sequence stars such as the Sun \cite{eddington}. For a polytropic equation of 
state, one can straightforwardly solve the continuity equation, 
Eq. (\ref{cont}):
\be{polysol}
\rho(r)=\kappa^{1/(1-\gamma)}\Big((\frac{B(r)}{\bar B})^{(1-\gamma)/(2\gamma)}-
1\Big)^{1/(\gamma-1)}.
\ee
Requiring, that $\rho$ vanishes at the stellar surface $r=r_R$, where 
$\bar B = B(r_R)$, sets $\gamma>1$. (It can be shown that requiring finite radius 
constrains $\gamma> 6/5$ in the Newtonian Lane-Emden -model \cite{weinberg2}.) 
Similarly, since $\rho'$ is also vanishing at the boundary \cite{Multamaki3}, we 
must further require that $\gamma<2$. 

For numerical work, it is advantageous to use scaled variables, $\rho=\rho_0 
\theta^{1/(\gamma-1)}$, 
$r=\alpha x$, $\alpha=\sqrt{\kappa\gamma/(4\pi G(\gamma-1))}\rho_0^{(\gamma-2)/2}$ 
(see eg. \cite{weinberg2}). In these variables, the fundamental equation of a 
Newtonian star reduces to the Lane-Emden -equation. In the case of GR or modified 
gravity, this is not the case, but the same change of variables is still useful. 
Using $\theta$ instead of $\rho$ is advantageous also due to the fact that unlike 
$\rho'$, $\theta'$ does not vanish at the boundary of the star, making 
identification of the star's surface easier.

\section{Numerical results}

Our next task is to compute numerically stellar profiles for certain polytropic 
cases for the $f(R)=R-\mu^4/R$ model. We have done this for a number of polytropic 
equations of state, both starting from the center of the star and from the 
boundary. The computed profiles $\theta$ as well as metric components, $A$ and $n$,
can then be compared to corresponding functions of Newtonian polytropes determined 
by the Lane-Emden -equation. Note that we require that stellar solutions have 
finite radii, unlike \eg toy models where the profile is approximated by an 
exponential function.

We consider the Sun, as a representative of main sequence stars, white dwarfs as 
well as neutron stars. The polytropic model of the Sun, with 
$\gamma_{\odot}=1.2985$, $\rho_0=1.53\times 10^5\,{\rm kg\, m}^{-3}$ and 
$p_0=3.00\times 10^{16}\, {\rm N\,m}^{-4}$ gives a fair approximation to the 
Stellar Standard Model \cite{Hendry}. For the relativistic white dwarfs, we use 
$\gamma=4/3,\ \kappa=1.4\times 10^{-7}({\rm kg\,m}^{-3})^{-1/3}$ and for 
nonrelativistic neutron stars
$\gamma=5/3,\ \kappa=3.5\times 10^{-11}({\rm kg\,m^{-3}})^{-2/3}$\cite{weinberg2}.

\subsection{Solution starting from the center outwards}

\begin{figure}[ht]
\begin{center}
\includegraphics[width=70mm,angle=0]{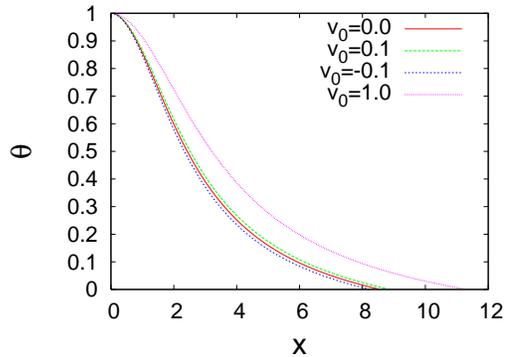}
\caption{Density profiles of the polytropic model of the Sun for 
different values of $v_0$.}\label{sun}
\end{center}
\end{figure}

\begin{figure}[ht]
\begin{center}
\includegraphics[width=70mm,angle=0]{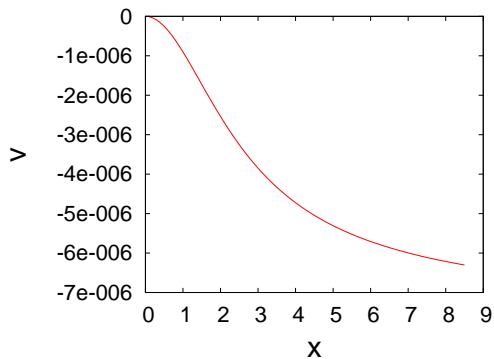}
\caption{Evolution of $v(r)=F(r)-4/3$ for the Sun, $v_0=0,\ r=\alpha x$.}\label{vb0}
\end{center}
\end{figure}

Using Eqs (\ref{origsol}) as a starting point, we can now integrate numerically 
the field equations along with the structural equation, Eq. (\ref{struct}), for a 
given stellar model. Integration is stopped at the stars surface \ie when 
$\theta=0$. Fixing the central density, $\rho_0$, fixes also $p_0$ via the 
polytropic equation of state but $v_0$ remains as a free parameter. In Fig. 
\ref{sun} we show the density profile for the Sun for a range of values of $v_0$. 
We see how changing $v_0$ scales the profile so that a larger value of $v_0$ leads 
to a star with a larger radius and vice versa. The corresponding 
Lane-Emden -solution \ie the polytropic model of the Sun is equal to choosing $v_0=0$ 
with very high precision.
In Fig. \ref{vb0} we show the evolution of $v$ as a function of the scaled radial 
distance $x$. Two properties are notable: the evolution is very small \ie the 
value changes very little over the radius of the star and the value is 
monotonically decreasing. The latter property is important when we consider 
fitting the star to a SdS-spacetime ($v=0$) outside the star.

Although the density profile in general closely resembles the Newtonian one, the 
behaviour of the metric is completely changed. Fitting the general metric to the 
PPN SdS-solution,
\bea{PPMm}
B(r)&=& 1-\frac {2 G M}{r}- H^2 r^2\nonumber \\ 
A(r)^{-1}&=& 1-\gamma_{PPN} \frac {2 G M}{r}- H^2 r^2,
\eea
we can solve the PPM-parameter $\gamma_{PPN}$ at distance $r$ as
\be{gamma}
\gamma_{PPN}= \frac{1+\frac{B(r)}{r B'(r)}}{1-\frac{A(r)}{rA'(r)}}.
\ee
Note that here we have neglected the $H^2 r^2$ -term as the cosmological parameter 
is extremely tiny compared to relevant solar system scales.

\begin{figure}[ht]
\begin{center}
\includegraphics[width=70mm,angle=0]{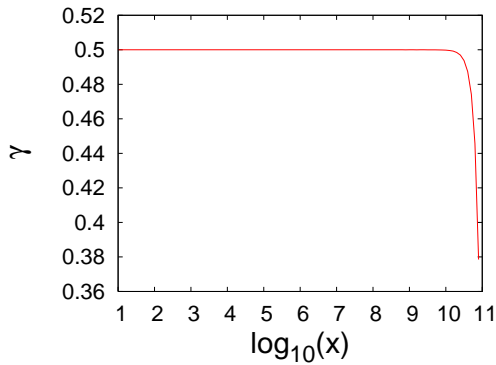}
\caption{Evolution of the PPN-parameter $\gamma_{PPN}$ outside the Sun.}
\label{ppnfig}
\end{center}
\end{figure}

Numerical work shows, that at the stellar boundary $r_R$ $\gamma_{PPN}$ tends to be 
near value $\gamma_{PPN}=0.5$, with small variations depending on the value $v_0$. 
In Fig. \ref{ppnfig} we show the evolution of $\gamma_{PPN}$ outside the Sun for 
the $v_0=0$ case but other choices of $v_0$ give essentially identical results.
The corresponding physical distance is equal to 
$r\approx x\times 8\times 10^8\,{\rm m}$.
From the figure we see that the spacetime outside the Sun in the CDTT-model is in 
gross violation of the experiments, remaining very close to $1/2$ far outside the 
Solar System. The behaviour at very large $x$ is explained by the fact that we 
ignore the cosmological term in the calculation of $\gamma_{PPN}$, Eq. 
(\ref{gamma}). The effect of the cosmological term should be included when
$2GM/r\sim H_0^2r^2\sim\mu^2 r^2$, \ie  $x\sim 10^{11}$, in good agreement 
with the numerical calculation. 

Note that the scale $l_{PPN}$ at which we expect the value of $\gamma$ to approach 
unity, \ie the range of the effective scalar in corresponding scalar-tensor theory
(see \eg \cite{odintsov,damour,magnano} and reference therein), is 
$l_{PPN}\sim 1/(\alpha\mu)\sim 10^{18}$ which far beyond the scale at which the 
cosmological term becomes effective. In the context of this particular $f(R)$ 
model, if one wishes to have $\gamma_{PPN}\to 1$ on scales where we can ignore the 
cosmological term, \ie $2GM/r\gg \mu^2r^2$ when $r\sim 1/\mu$, sets 
$\mu\gsim 1/{\rm km}$ for the Sun, $M\approx 2\times 10^{30}\,{\rm kg}$. This shows 
how in this model we cannot choose the different relevant scales to have 
physically meaningful values.

We have performed corresponding calculations also for relativistic white dwarfs and
nonrelativistic neutron stars. Although stellar numerical values and scales are 
clearly different, qualitative conclusions remain: density profiles are close to
Newtonian ones whereas metric profiles are completely different and 
$\gamma_{PPN}\approx 0.5$. We have also considered configurations with pressure
comparable to density
in which case we find that $\gamma_{PPN}$ can deviate significantly from $1/2$.

\subsection{Solutions with the SdS metric as the external solution}

We have seen that starting with a regular solution at the origin leads to 
unacceptable space-time outside the star, in good agreement with previous work 
\cite{Erickcek:2006vf,Chiba2,Jin:2006if, Faulkner:2006ub, Kainulainen2007}. This 
result suggests that starting with a physically acceptable outside solution, the 
SdS-metric, will lead to an irregular solution at the center of the star.

\begin{figure}[ht]
\begin{center}
\subfigure[\ Density profiles.]{\includegraphics[width=70mm,angle=0]{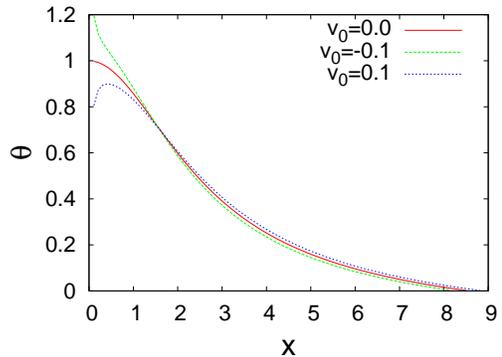}}
\subfigure[\ Profiles of $v(r),\ r=\alpha x$.]{\includegraphics[width=70mm,angle=0]{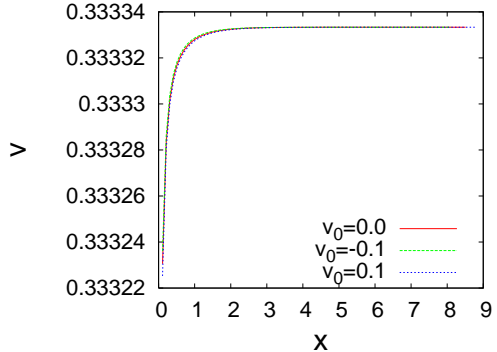}}
\caption{Stellar configurations with external SdS-metric.}\label{outin}
\end{center}
\end{figure}

In order to determine the relevance of the SdS-metric for the particular 
$f(R)$ model in question, we solve Eqs (\ref{numequs}) starting from the stellar 
boundary at given radius $r_R$. We set the boundary conditions by requiring that 
external metric is the SdS-metric
\be{SdS}
B_{ext}(r)=A_{ext}(r)^{-1}=1-\frac{2 G M}{r}-H^2 r^2,
\ee
where $H^2=\mu^2/(4\sqrt{3})$. At the boundary we require that 
$n(r_R)=B'_{ext}(r_R)/B_{ext}(r_R),\ A(r_R) = A_{ext}(r_R)$ as well as 
$v(r_R)=v'(r_R)=\theta (r_R)=0$, corresponding to the SdS-metric \cite{Multamaki3}.
In practice we first fix central density $\rho(0)$, and then set the gravitating 
mass $M$ and stellar radius $r_R$, by solving equations from the center outwards 
for a given $v_0$. The computed mass and radius are then used as parameters in the 
outside metric, which then fixes the boundary condition for $n$ and $A$. 
In Fig. \ref{outin}(a), we show the stellar profiles corresponding to different 
choices of $v_0$ \ie corresponding to different values of $M$ and $r_R$. From the 
figure we see that for $v_0=0$, or the Sun, the density profile is unchanged 
\ie fixing the outside solution to the SdS-metric leads to a physically acceptable 
density profile. When $v_0$ deviates from zero, the solution diverges near the 
origin. 
In the same figure we also show the corresponding evolution of $v$, Fig. 
\ref{outin}(b). In all of the cases the conclusion is the same: $v$ diverges at the 
origin. This is as expected since we have already seen that any solution with a 
regular metric and density profile at the origin leads to a experimentally 
unacceptable outside solution.

Requiring the SdS-solution as the outside metric leads to a divergent $v$ and
hence the scalar curvature diverges at the origin. Similarly also the metric 
components $A$ and $n$, are irregular at the origin. In summary, if one relaxes the 
requirement of regularity of the metric at the origin, one can have the SdS-metric as 
the outside solution and reproduce the density profile.

It is worth noting, that if the star does not follow polytropic equation of state, 
especially in the core of the star, non-singular solution with external SdS-metrics 
can be found when $v_0=0$. In particular the regularity of the scalar curvature is 
seen directly from Eq. (\ref{econtra}) if the equation of state near the center of 
the star is relativistic $p=\rho/3$. We have confirmed this phenomenon by using an 
equation of state, which is polytropic in the outer region and relativistic in the 
core. This resembles the case of massive, relativistic neutron stars \cite{weinberg2}.

\section{Comparison with General Relativity}

We have seen that the case $v_0=0$ reproduces the result from using the 
Lane-Emden -equation very well. We can understand this behaviour analytically by 
considering the field equations in GR and in the f(R) theory.

In GR, the field equations can be written in the form 
$R_{\mu}^\nu=8\pi G(T_\mu^\nu-g_\mu^\nu\, T/2)$, where $T=T_\mu^\mu$.
In the limit of negligible pressure, the $00$-component reads as 
$R_0^0\approx 4\pi G\rho$. The scalar curvature, $R$, follows the density 
\ie $R\sim\rho$.

\begin{figure}[ht]
\begin{center}
\psfrag{eqK}{$\frac{3\Box\hskip-0.7pt F}{8\pi G\rho}$}
\psfrag{R00}{$\frac{R_0^0 F}{\Box\hskip-0.7pt F}$}
\includegraphics[width=70mm,angle=0]{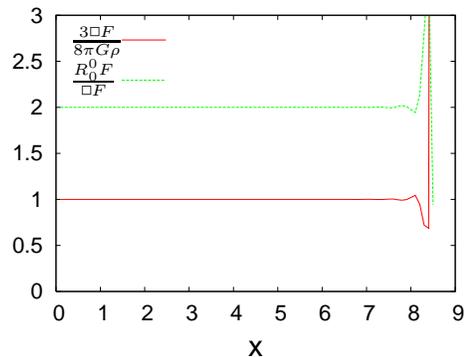}
\caption{Comparison of the size of different terms of the field equations for 
the Sun.}
\label{Kcomp}
\end{center}
\end{figure}

In the CDTT-model, the situation is different. Now, as numerical calculations show, 
inside and outside the star, $f(R)\sim R\sim\mu^2$ and $F\sim \cO(1)$. From the field 
equations one would then expect that for the $00$-component, the $FR_0^0$ and $\Box F$ 
terms to be dominant \ie $FR_0^0+\Box F\approx 8\pi G \rho$ or if one considers
the contracted equation, Eq. (\ref{econtra}), $3\Box F\approx 8\pi G\rho$.
In Fig. \ref{Kcomp} we plot the relevant terms for the
Sun (here $v_0=0$, but the situation is unchanged if $v_0$ is varied as long as 
$F\gg f$). From the figure we see that inside the star the approximations hold and 
hence, we can write 
$F R_0^0\approx 16\pi G\rho/3$. From the numerical work we furthermore know, that
 the value of $v$ or $F$ changes very little inside the star and hence we can 
approximate $F\approx const.=F_0=4/3+v_0$. We have then
\be{Rapp}
R_0^0\approx  \frac{16\pi G}{3 F_0}\rho\approx \frac{4\pi G}{1+\frac{3}{4}v_0}\rho.
\ee

\begin{figure}[ht]
\begin{center}
\includegraphics[width=70mm,angle=0]{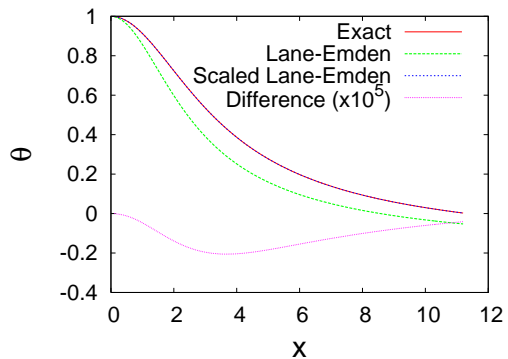}
\caption{Comparison of exact $f(R)$-model, Lane Emden and scaled Lane-Emden -model
density profiles.}\label{comparison}
\end{center}
\end{figure}

Comparing to the GR expression, we see that there exists a simple scaling between the
two expressions. In the Lane-Emden -equation this scaling signals its presence in 
the scaling of the radial coordinate, $r=\alpha x\propto x/\sqrt{G}$, if $\rho_0$ is 
fixed \ie we expect that 
\be{scaling}
\theta_{f(R)}(x)\approx \theta_{GR}(\frac{x}{\sqrt{1+\frac{3}{4} v_0}}).
\ee
In Fig. \ref{comparison} we show the Sun's profile with $v_0=1$ (solid red line) along 
with the Lane-Emden -solution $\theta_{LE}(x)$ (dotted green line) and a  scaled 
Lane-Emden -solution, $\theta_{LE}(x/\sqrt{1+(3/4)v_0})$. The line at the bottom is the 
difference between the scaled Lane-Emden -solution and the numerical solution for 
$v_0=1$ magnified by a factor of $10^5$. As we can see, the scaled solution reproduces 
the numerical solution very well.

One can hypothesize that a similar scaling would exist in a general $f(R)$ theory, 
$f(R)=R+g(R,\mu)$. In order for such a theory to explain late time acceleration, the 
extra terms will in general have a new scale $\mu$ associated with the value of the 
present Hubble parameter (in principle one could also consider very finely tuned 
theories, where different parameters with different scales would conspire to produce 
current acceleration). The vacuum state of the theory is such that it possesses 
non-zero constant curvature, $R_0\sim \mu^2$. Examples of such theories are 
$f(R)=R-\mu^4/R$ and $f(R)=R-\mu^4/R+R^2/\beta^2$. In the latter case, choosing 
$\beta\sim \mu$ helps to avoid Solar System constraints \cite{Chiba2} by 
making the mass of effective scalar large. If $R$ remains close to the cosmological 
value inside the star, like in the CDTT-model, it is natural to expect that 
$f\sim R\sim\mu^2\sim H_0^2$ and $F\sim \cO(1)$ and hence a similar scaling property
should apply.

\section{Conclusions}

In the present paper we have analyzed the properties of polytropic stars in a
generalized gravity model. In particular we have considered the $f(R)$ model with 
$f(R)= R-\mu^4/R$ with the conclusion that the density profiles in general resemble 
the Newtonian Lane-Emden -solutions. Requiring that stellar solution is regular at 
the origin, we found that slightly varying the central curvature, $v_0$, the stellar 
mass and radius are changed but preserve their functional Lane-Emden -form. However,
the metric components are drastically different from the Lane-Emden -case and 
therefore our results for the external metric conform to previously calculated results 
for completely pressure-less matter \cite{Chiba2,Kainulainen2007}. In 
particular the PPN parameter $\gamma_{PPN}$ outside the start is near 
$\gamma_{PPN} = 1/2$. 
 
If we do not require complete regularity of stellar solution at the origin, but 
assume the external SdS-solution, we still find stellar profiles in good agreement with 
the Lane-Emden solution. Differences appear only near the center of the star and
these deviations depend on how much the mass and radius differ from the corresponding
Newtonian configuration. The interior solution for metric components and curvature 
for such stars are always singular, although stars with relativistic matter
at the core may evade this property.

Consequently, the $f(R)= R-\mu^4/R$ model is not experimentally suitable to describe
the space time around the Sun. A possible way out is to relax the requirements set 
for the central boundary conditions, but a more plausible approach is to modify the 
functional form of $f(R)$. The form of the action function $f(R)$ should
differentiate the cosmological Hubble scale $R_0 \sim H_0^2$ determined by
$2 f(R_0)=R_0 f'(R_0)$ and the effective scalar mass scale $\propto 1/f''(R_0)$.
Then it may be possible to have $\gamma_{PPN} \goto 1$ at a distance $l_{PPN}$ small 
compared to solar system distances, possibly redeeming some of the $f(R)$ models. 
This is, however, not possible in the $f(R)= R-\mu^4/R$ model, because both scales 
are controlled by a single parameter $\mu$, which when set to the cosmologically 
relevant value $\mu\sim H_0$, leads to $l_{PPN}\gg 1/H_0$. In other words, the 
asymptotic SdS-metric is never reached. 

Seeking possible ways to save $f(R)$ models note also, that as discussed in 
\cite{Multamaki3}, a suitable choice of $f(R)$ may exactly reproduce GR/Newtonian 
density profiles changing only $A$ and $v$. In this case we do not, however, know
much about their behaviour neither outside the star nor near the center of the star 
and even cosmological constraints are unknown.  

Possibly several more general $f(R)$ models are physically acceptable, but in 
particular the model with $f(R)=R-\mu^4/R+R^2/\beta^2$ may do after considerable 
fine-tuning of $\beta$. Note however, that in this case an important 
distinction compared to the CDTT-model applies. In the CDTT-model, $F_0\approx 4/3$ or $v_0\approx 0$,
inside and outside the star so that the field equations are effectively similar to the
GR counteparts, Eq. (\ref{Rapp}). When the $R^2$ term is added to the action,
$F_0\neq 4/3$ outside the star and hence if we still wish to have $F_0\approx 4/3$ inside star
in order to reproduce the Lane-Emden -solution, $F$ must evolve significantly over the radius
of the star. Otherwise, if $F_0$ remains approximately constant, either the SdS-solution is not
the correct outside solution, or the radius of the star will be different than in GR. 
This argument potentially offers a new, general constraint on $f(R)$ theories of gravity, 
motivating further work.

\acknowledgments

We thank Kimmo Kainulainen, Vappu Reijonen and Daniel Sunhede as well as Chris Flynn 
for valuable discussions during this work. KH is supported by the Wihuri foundation.
TM gratefully acknowledges support from the Academy of Finland.



\end{document}